\documentclass[a4paper,fleqn,usenatbib,useAMS]{mnras}

\usepackage{graphicx}	
\usepackage{amsmath}	
\usepackage{amssymb}	
\usepackage{multicol}        
\usepackage{bm}		
\usepackage{pdflscape}	

\usepackage[T1]{fontenc}
\usepackage{ae,aecompl}

\usepackage{newtxtext,newtxmath}

\title[Tracing the Sources of Reionization]{Tracing the Sources of Reionization in Cosmological Radiation Hydrodynamics Simulations}

\author[H. Katz]{Harley Katz$^{1}$\thanks{Contact e-mail: \href{mailto:harley.katz@physics.ox.ac.uk}{harley.katz@physics.ox.ac.uk}}, Taysun Kimm$^{2}$, Martin G. Haehnelt$^3$, Debora Sijacki$^3$, Joakim Rosdahl$^4$, \newauthor and Jeremy Blaizot$^4$
\\
$^{1}$Astrophysics, University of Oxford, Denys Wilkinson Building, Keble Road, Oxford OX1 3RH, UK\\
$^{2}$Department of Astronomy, Yonsei University, 50 Yonsei-ro, Seodaemun-gu, Seoul 03722, Republic of Korea\\
$^{3}$Institute of Astronomy and Kavli Institute for Cosmology, Cambridge, Madingly Road, Cambridge, CB3 0HA, UK\\
$^{4}$Univ Lyon, Univ Lyon1, ENS de Lyon, CNRS, Centre de Recherche Astrophysique de Lyon UMR5574,
F-69230, Saint-Genis-Laval, France}

\date{\today}

\pubyear{2017}

\begin{document}
\label{firstpage}
\pagerange{\pageref{firstpage}--\pageref{lastpage}}
\maketitle

\begin{abstract}
We use the photon flux and absorption tracer algorithm presented in Katz~et~al.~2018, to characterise the contribution of haloes of different mass and stars of different age and metallicity to the reionization of the Universe. We employ a suite of cosmological multifrequency radiation hydrodynamics AMR simulations that are carefully calibrated to reproduce a realistic reionization history and galaxy properties at $z \geq 6$. In our simulations, haloes with mass $10^9{\rm M_{\odot}}h^{-1}<M<10^{10}{\rm M_{\odot}}h^{-1}$, stars with metallicity $10^{-3}Z_{\odot}<Z<10^{-1.5}Z_{\odot}$, and stars with age $3\,{\rm Myr} < t < 10 \, {\rm Myr}$ dominate reionization by both mass and volume.  We show that the sources that reionize most of the volume of the Universe by $z=6$ are not necessarily the same sources that dominate the meta-galactic UV background at the same redshift.  We further show that in our simulations, the contribution of each type of source to reionization is not uniform across different gas phases. The IGM, CGM, filaments, ISM, and rarefied supernova heated gas have all been photoionized by different classes of sources.  Collisional ionisation contributes at both the lowest and highest densities.  In the early stages of the formation of individual HII bubbles, reionization proceeds with the formation of concentric shells of gas ionised by different classes of sources, leading to large temperature variations as a function of galacto-centric radius. The temperature structure of individual HII bubbles may thus give insight into the star formation history of the galaxies acting as the first ionising sources. Our explorative simulations highlight how the complex nature of reionization can be better understood by using our photon tracer algorithm. 
 \end{abstract}




\section{Introduction}
Identifying and categorising the sources that reionized the Universe remains one of the primary goals of high-redshift extragalactic astrophysics.  The launch of JWST in 2019 will surely reduce some of the uncertainty surrounding this question \citep[e.g.,][]{Windhorst2006}; however, currently, no fewer than eight classes of sources have been proposed as contributors to reionization.  These include dwarf galaxies \citep{Couchman1986}, massive galaxies \citep{Sharma2016}, active galactic nuclei \citep{Madau2015,Haiman1998,Madau1999}, accretion shocks \citep{Dopita2011}, globular clusters \citep{Ricotti2002,Katz2013,Katz2014}, stellar mass black holes \citep{Madau2004,Ricotti2004,Mirabel2011}, and dark matter annihilation and decay \citep{Mapelli2006}.  

The timescale and physics of reionization is strongly dependent on the sources that ionise the Universe.  In a galaxy dominated reionization scenario, the first ionising photons are emitted from the highest density regions of the Universe.  Galaxy formation is strongly clustered and thus the highest sigma peaks in the matter distribution may be ionised before the lowest density IGM leading to an inside-out reionization scenario \citep[e.g.,][]{Furlanetto2004,Iliev2006,Lee2008,Friedrich2011}.  The highest density regions of the Universe however have the shortest recombination rates.  Thus if the early emitted ionising photons can penetrate out to the lowest density regions of the IGM, the lowest density gas may indeed remain ionised while the highest density regions return to a neutral state \citep[e.g.,][]{ME2000,Choudhury2009}.  The presence of filaments complicates this picture since they have a much higher recombination rate compared to the surrounding IGM and thus it has been postulated that these are the last regions of the Universe to reionize \citep[e.g.,][]{Gnedin2000a,Finlator2009,Katz2017}.  Because the different types of sources control the topology of reionization and the sizes of HII bubbles, the shape and amplitude of the 21cm signal is strongly affected \citep[e.g.,][]{Zaldarriaga2004,McQuinn2007b,Kulkarni2017}.  Furthermore, the amount of heating in the IGM is dependent on the spectral energy distribution (SED) of the source and this will also impact the 21cm signal \citep[e.g.,][]{Madau1997}.

In the context of galaxy formation, photoheating due to reionization can have a strong effect in suppressing the formation of dwarf galaxies \citep[e.g.,][]{Gnedin2000,Okamoto2008,Gnedin2014,Xu2016b,Dawoodbhoy2018}.  The exact mass scale where this occurs is debated; however, recent high-resolution simulations demonstrate that most systems with virial mass at or below the atomic cooling threshold suffer some degree of inflow suppression and can also be photoevaporated by the UV background.  Because reionization is inhomogeneous, the properties of dwarf galaxies may change depending on environment and the topology of reionization \citep{Aubert2018,Dawoodbhoy2018}.  

When attempting to calculate the contribution of different classes of sources to reionization, it is commonplace to measure the escape fraction of photons at the virial radius of a galaxy \citep[e.g.,][]{Kimm2017,Trebitsch2017,Rosdahl2018}.  Using a simple 1D equation \citep[e.g.][]{Robertson2015} that combines the escape fraction with a stellar IMF and a star formation rate constrained from observations, the contribution of various sources to reionization can be estimated.  However, measuring only the escape fraction does not take into account the environmental dependence of the number of photons that reach the low density IGM.  To address this, \cite{Katz2018} developed a photon tracer and absorption algorithm to directly measure the contribution of any arbitrary class of sources to the meta-galactic UV background and the reionization of gas, both spatially and temporally, as a function of redshift.  In that work, both the reionization history and galaxy properties were carefully calibrated and the contribution of different mass haloes, different metallicity stars, and different age stars to the HI photoionisation rate was calculated as a function of redshift.  

In this work, we use our photon tracer and absorption algorithm to measure the contribution of three different classes of sources depending on halo mass, stellar metallicity, or stellar age to the reionization of the Universe by volume, mass, and environment.  While in \cite{Katz2018} we studied how these different classes of sources contributed to the UV background,  here, {\em we concentrate on how photons emitted by different classes of sources interact with the gas and drive reionization}.

The paper is organised as follows. In Section~\ref{simdis}, we briefly review the setup of our cosmological multifrequency radiation hydrodynamics simulations that were presented in \cite{Katz2018}.  In Section~\ref{results}, we demonstrate how the growth of HII bubbles and their temperature structure is inherently linked with the bursty star formation rate of the galaxy and the types of sources embedded in the interstellar medium (ISM) and show how haloes of different mass, and stars with different ages and metallicities contribute to reionization by mass, volume, and environment.  In Section~\ref{cavs}, we discuss caveats associated with this generation of simulations.  Finally, in Section~\ref{discussion}, we present our discussion and conclusions and highlight how the photon tracer and absorption algorithm can be exploited in future work.

\section{Simulations}
\label{simdis}
Our analysis here is based on the simulations presented in \cite{Katz2018}.  Briefly, initial conditions have been generated in a 10cMpc$h^{-1}$ box at $z=150$ with a uniform grid of $256^3$ dark matter particles ($m_{\rm dm}=6.51\times10^6{\rm M_{\odot}}$) using {\small MUSIC} \citep{Hahn2011}.  Our cosmological parameters have been adopted from \cite{Planck2016} such that $h=0.6731$, $\Omega_{\rm m}=0.315$, $\Omega_{\Lambda}=0.685$, $\Omega_{\rm b}=0.049$, $\sigma_8=0.829$, and $n_s=0.9655$.  All of the gas in the box is initially neutral and is composed of 76\% hydrogen and 24\% helium by mass. 

\begin{table}
\centering
\begin{tabular}{@{}llll@{}}
\hline
Bin & E$_{\text{min}}$ & E$_{\text{max}}$ & Function\\
 & [eV] & [eV] & \\
\hline
1 & 13.60 & 24.59 & H Photoionisation\\
2 & 24.59 & 54.42 & He \& H Photoionisation\\
3 & 54.42 & $\infty$ & He, He$^+$, \& H Photoionisation\\
\hline
\end{tabular}
\caption{Photon energy bins used in the simulation.}
\label{ebins}
\end{table}

The simulations were run with the publicly available multifrequency radiation hydrodynamics code  {\small RAMSES-RT} \citep{Rosdahl2013} which is an extension of the adaptive mesh refinement (AMR) code {\small RAMSES} \citep{Teyssier2002}.  In our current work, we use three frequency bins, all in the UV, listed in Table~\ref{ebins}, that are coupled to a six species non-equilibrium chemistry model that follows H, H$^+$, $e^-$, He, He$^+$, and He$^{++}$.  The gas is allowed to cool via collisional ionisations, recombinations, collisional excitation, Bremsstrahlung, Compton cooling (and heating), and dielectronic recombination for H and He and their ions.  Metal line cooling is modelled at $T>10^4$K using interpolated tables from {\small CLOUDY} \citep{Ferland2013} and at $T<10^4$K using analytic fits to the cooling function adopted from \cite{Rosen1995} based on \cite{Dalgarno1972}.

\begin{table}
\centering
\begin{tabular}{@{}llll@{}}
\hline
Halo Mass & Colour  & Line Style & Range  \\
 & in Plots & & \\
\hline
T1 & {\color{red} Red} & Dashed & $M<10^9{\rm M_{\odot}}h^{-1}$ \\
T2 & {\color{green} Green} & Dot-Dashed & $10^9{\rm M_{\odot}}h^{-1}<M<10^{10}{\rm M_{\odot}}h^{-1}$ \\
T3 & {\color{blue} Blue} & Dotted &$M>10^{10}{\rm M_{\odot}}h^{-1}$ \\
\hline
Stellar  &  & & \\
Metallicity & & & \\
\hline
T1 & {\color{red} Red} & Dashed & $Z<10^{-3}Z_{\odot}$ \\
T2 & {\color{green} Green} & Dot-Dashed & $10^{-3}Z_{\odot}<Z<10^{-1.5}Z_{\odot}$  \\
T3 & {\color{blue} Blue} & Dotted & $Z>10^{-1.5}Z_{\odot}$\\
\hline
Stellar Age & &  &\\
\hline
T1 & {\color{red} Red} & Dashed & $t < 3 \, {\rm Myr}$ \\
T2 & {\color{green} Green} &  Dot-Dashed & $3\,{\rm Myr} < t < 10 \, {\rm Myr}$ \\
T3 & {\color{blue} Blue} & Dotted & $t > 10 \, {\rm Myr}$\\
\hline
\end{tabular}
\caption{Photon tracer bins used in the three fiducial simulations.  The most massive halo has mass $M{\rm=1.2\times10^{11}M_{\odot}}$ and the most metal enriched star has metallicity $Z=10^{-1.2}Z_{\odot}$.}
\label{photbins}
\end{table}

\begin{figure*}
\centerline{\includegraphics[scale=1,trim={0 0 0 0},clip]{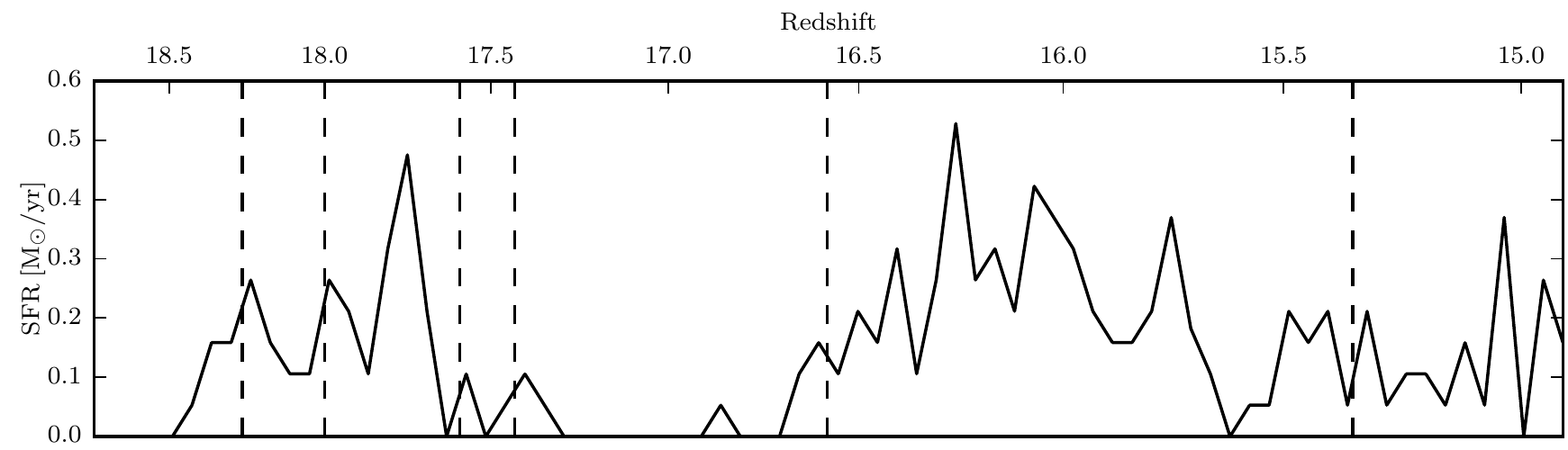}}
\centerline{\includegraphics[scale=1,trim={0 0 0 0},clip]{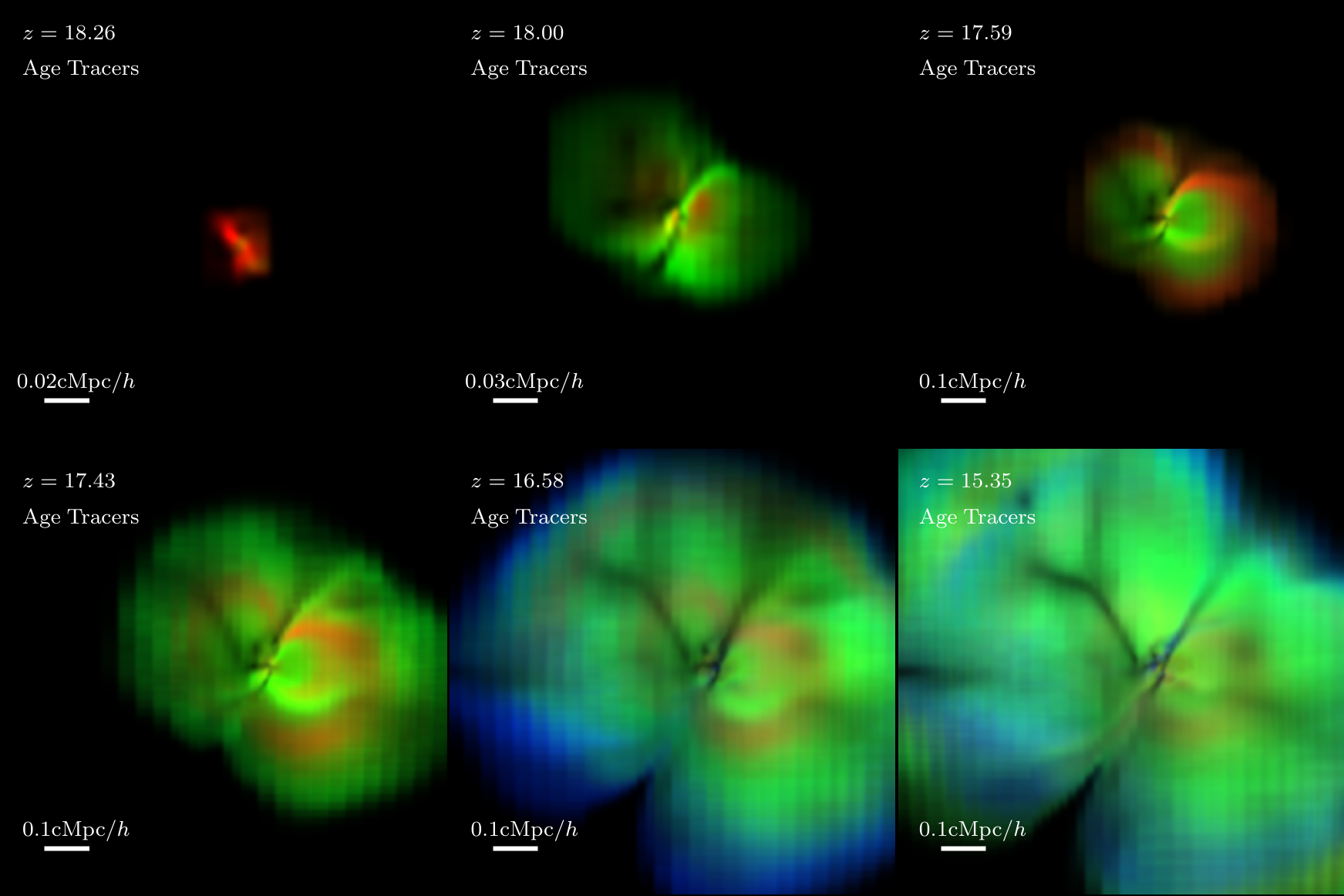}}
\caption{{\it Top}.  Star formation rate averaged over 1Myr intervals as a function of redshift for the most massive halo in the simulation in the redshift range $15<z<18.5$.  The dashed vertical lines show the redshifts plotted in the bottom figure.  {\it Bottom}. The evolution of the first HII bubble when tracing by stellar age around the most massive galaxy in the simulation.  The intensity represents the mass-weighted HII fraction for each tracer along a depth of $1$cMpc$h^{-1}$.  Red, green, and blue represent young (T1), middle aged (T2), and old (T3) stars, respectively.  As redshift decreases, subsequent generations of stars are formed and ionising photons are injected into the IGM increasing the size of the HII bubble.  As the stellar populations age, multiple concentric shells are formed, providing a record of which generation of stars ionised the gas.  When the denser regions of the bubble begin to recombine, the colours begin to mix as the gas is partially ionised by different generations of stars.}
\label{onion}
\end{figure*}

Star formation occurs in cells with $n_{\rm H}\geq1$cm$^{-1}$ following a Schmidt law \citep{Schmidt1959} with an efficiency of 1\% per free-fall time.  10Myr after formation, star particles explode as Type~II supernovae, and mass, metals, and thermal energy are injected into the gas as described in \cite{Katz2017}.  In order to combat numerical overcooling \citep[e.g.][]{Katz1992}, we use the delayed cooling supernova feedback model of \cite{Teyssier2013}, where the cooling is shut off in cells affected by a supernova for 20Myr.  Star particles also inject ionising photons into the gas over time as a function of their mass, metallicity, and age using the BPASSv2 SED \citep{BPASS,Stanway2016}.  We have adopted the SED model with a maximum stellar mass of 300M$_{\odot}$ and an IMF slope of $-2.35$ between stellar masses of $0.5<M/{\rm M_{\odot}}<300$ and a slope of $-1.35$ between stellar masses of $0.1<M/{\rm M_{\odot}}<0.5$.  In order to reproduce a realistic reionization history, we have scaled the normalisation of the SED by a factor of 1.193, thus increasing the total luminosity of the SED by a small amount.

\begin{figure*}
\centerline{\includegraphics[scale=1,trim={0 0 0 0},clip]{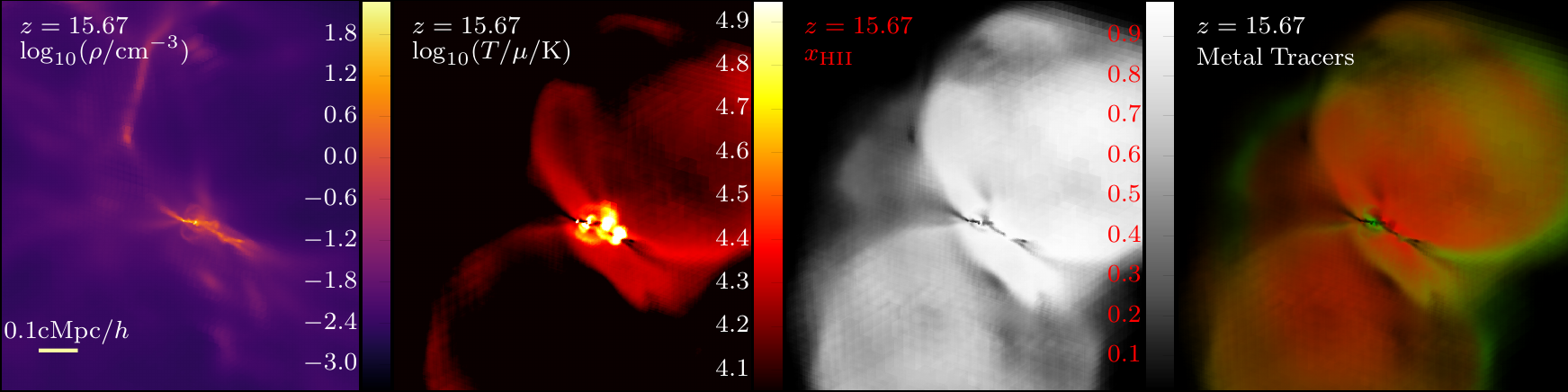}}
\caption{Snapshot of density, temperature divided by the mean molecular weight ($\mu$), HII fraction ($x_{\rm HII}$), and HII fraction coloured by the stellar metallicity tracers at $z=15.67$ in a $1$cMpc$h^{-1}$ cube around the most massive galaxy in the simulation.  Quantities represent the mass-weighted mean along the axis of projection.  The HII bubble is centred around the densest region and exhibits a double lobed structure around the filaments that are feeding the galaxy.  The bubble does not have a uniform ionisation fraction as can be seen in the third panel and is exhibited in the temperature structure of the system.  In the fourth panel, a shell structure has emerged whereby the inner regions have been ionised by the lowest metallicity stars (T1) while the thin green outer shell has been ionised by intermediate metallicity stars (T2).  The differences in the temperature structure can be related to the sources that ionised the gas as the hotter regions were correspondingly ionised by the intermediate metallicity stars.}
\label{metal}
\end{figure*}

Refinement occurs on-the-fly when a cell contains eight times its mass in either dark matter or baryons as initialised on the coarse grid.  The simulation refines up to level 14 and attempts to maintain a constant physical resolution of 125pc.  In order to properly model the propagation of ionisation fronts in both low and high density regions of the simulation, we exploit the variable speed of light approximation presented in \cite{Katz2017} and use an AMR level dependent speed of light.  We set the speed of light to be 0.1c on the base grid of the simulation and divide this by a factor of 2 on each subsequent level of refinement.  A convergence study on our choice of speed of light is shown in the appendix of \cite{Katz2018} where we find that using 0.1c on the coarse grid is adequate for the properties that we are interested in studying here.

Most importantly, these simulations have been calibrated to reproduce a realistic reionization history constrained by high-redshift observations \citep[e.g.][]{Fan2006}.  The simulations are also in good agreement with measurements of the HI photoionisation rate as a function of redshift \citep{Bolton2007,Wyithe2011,Calverley2011,Becker2013}.  Furthermore, the stellar feedback dissipation timescale was set so that the simulated systems fall on the stellar mass to halo mass relation predicted and extrapolated from abundance matching \citep{Behroozi2013}. However, one must keep in mind that our choice of stellar feedback and subgrid escape fraction are not a unique combination to reproduce the stellar mass to halo mass relation as well as the reionization history.  Different choices of subgrid models that require different parameters to match these observations may result in a change in the fractional contributions of each class of sources to the reionization history and the photoionization rate.  Due to the finite box size, we are not able to account for the contribution of haloes more massive than $\sim10^{11}$M$_{\odot}$ to reionization or the full range of densities and environments (e.g. deep voids) that would be present in larger boxes.  Likewise, our mass resolution limits our ability to probe the smallest dwarf galaxies with $M_{\rm vir}\lesssim10^{8.5}$M$_{\odot}$ (this contribution was estimated in Figure~4 of \citealt{Katz2018}).  Finally, the spatial resolution of our simulation is insufficient to resolve the ISM of galaxies and the escape of LyC photons from a multiphase ISM.  With these limitations in mind, we can nevertheless use these demonstrative simulations to better understand reionization, but caution that some conclusions may change when all of the relevant physical scales are resolved.  

We employ the same simulations using the photon tracer and absorption algorithm presented in Section~2 of \cite{Katz2018}.  Briefly, the photon tracer and absorption algorithm is able to separate the instantaneous contribution of any source population to the photoionization rate and the volume weighted ionization fraction, both spatially and temporally, by ``tagging" photons based on their source.  The algorithm works by separating the number density of photons in each cell into fractional contributions from $N_{\rm source}$ tracer groups where $N_{\rm source}$ represents the number of different source classes (e.g. Pop.~II stars vs. Pop.~III stars, vs. black holes).  These fractional contributions are then updated when photons are advected between cells or when star particles inject photons into their host cells.  Finally, when photons are absorbed in the non-equilibrium chemistry module, we calculate the fraction of HI in each cell that was ionized by photons from each of the different source populations.  These fractions are also stored and advected with the gas.  In total, we have run three simulations: in the first run, we trace the photons based on the mass of the halo from which they are emitted.  In the second run, we trace the photons based on stellar metallicity and in the third run, we apply our photon tracers to stellar age.  Table~\ref{photbins} lists the properties of each of our different tracer groups as well as the colour each tracer group is represented by in all of the figures.

\section{Results}
\label{results}

\subsection{How galaxies ionize their surroundings}
\label{gis}
We begin our analysis by demonstrating how photons are escaping from the galaxies in our simulations.  By combining the age tracers with the redshift evolution of the simulation, we can improve our understanding on how the growth and evolution of the HII bubbles relates to the star formation inside the galaxy and how this radiation leaks into the IGM.  In Figure~\ref{onion}, we show a time series of the 56Myr period from $z=18.26$ to $z=15.35$ that follows the formation of the first ionised bubble in the simulation along with the star formation rate of the halo.

\begin{figure*}
\centerline{\includegraphics[scale=1,trim={0 0 0 0},clip]{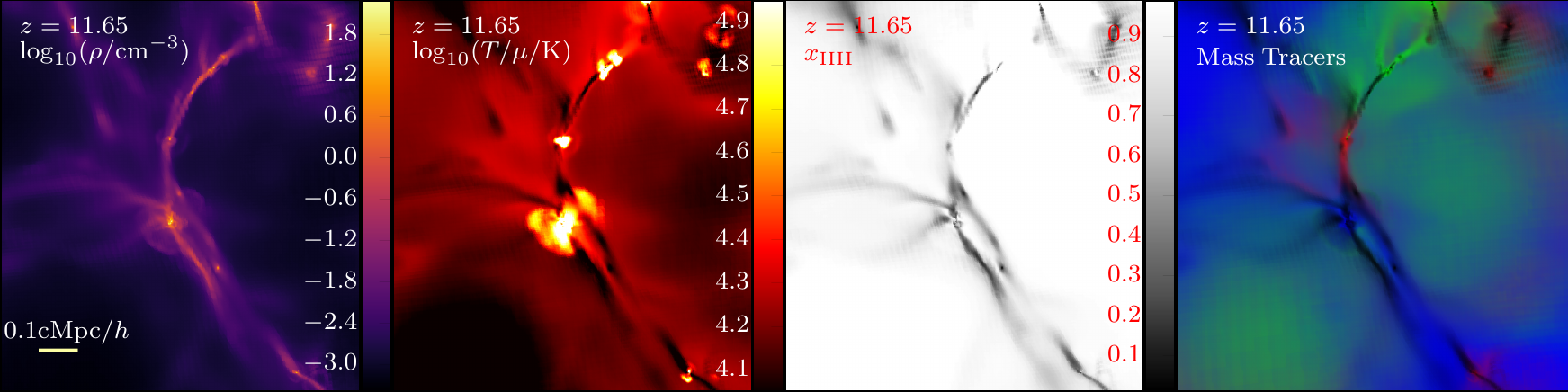}}
\caption{Snapshot of density, temperature, HII fraction, and HII fraction coloured by halo mass tracers at $z=11.65$ in a $1$cMpc$h^{-1}$ cube around the most massive galaxy in the simulation.  Quantities represent the mass-weighted mean along the projected axis.  Different regions surrounding the massive system are ionised by different mass galaxies.  In particular, the filaments are ionised by systems that form within the filaments.}
\label{mass_filament}
\end{figure*}

At $z=18.26$, the first generation of stars have just formed and have ages $<3\,{\rm Myr}$. A small, red HII bubble (indicating age $<3\,{\rm Myr}$) has formed around the galaxy.  In the $4\,{\rm Myr}$ period between $18.0<z<18.26$, the star formation rate has decreased and the first generation of stars has moved from the young star bin to the middle aged star bin.  Thus, the gas that these stars now ionise appears green.  The next snapshot at $z=18.0$, shows that a green HII region has formed outside of the galaxy, enveloping the previously ionised gas that still appears red.  The region is double lobed as two dense filaments are feeding the system approximately from the northeast and southwest, consistent with the ``butterfly" shapes of HII bubbles seen in other work \citep[e.g.][]{Abel1999,Ciardi2001}.  By volume, the ionisation of this HII bubble is dominated by the middle aged stars.  This is a result of combining two physical effects.  First, even though the young and middle aged stars emit similar cumulative amounts of ionising photons (see Figure~9 of \citealt{Katz2018} where for the lowest metallicity stars, 55\%, 35\%, and 9\% of the total number of photons are emitted by young, intermediate, and old age stars, respectively), clearly, more of the photons from the middle aged stars escape into the CGM.  Second, the regions further away from the system are less dense.  Thus the number of ionising photons needed to ionise the gas per unit volume decreases and therefore, the ionising photons emitted by stars can penetrate deeper into the IGM.  A shell structure has begun to form whereby a very small red HII bubble is embedded in a green HII bubble.

In the $7\,{\rm Myr}$ period between $17.59<z<18.0$, a large star formation event has occurred, as can be seen in star formation rate versus redshift plot at the top of Figure~\ref{onion}.  The luminosity of the system is dominated once again by the young stars.  However, ionised channels that lead to the IGM exist as a result of the first generation of stars and therefore, ionising photons from this new generation of young stars can penetrate out of the galaxy and into the IGM.  A thick red shell has now enveloped the green shell in the third panel of Figure~\ref{onion}.  Once this new generation of stars has aged, the newly ionised gas will once again appear green.  The fourth snapshot in Figure~\ref{onion} shows that, at $z=17.43$ ($2\,{\rm Myr}$ after the previous snapshot), a green shell has appeared outside of the newly created red shell.  There are now at least three distinct regions inside of this HII bubble (green, red, green) where the young and middle aged stars forming in the first and second generation bursts have impacted the gas.

In the $16\,{\rm Myr}$ period between the fourth and fifth snapshots ($16.58<z<17.43$) very little star formation occurs.  At $z=16.58$, a thin blue shell has formed at the edge of the HII bubble indicating that the ionising photons emitted by the stars older than $10\,{\rm Myr}$ are beginning to impact the IGM.  These photons only become visible when low enough densities are reached such that relatively few ionising photons ($<10\%$ of the total) emitted by these stars can ionise a significant volume of the IGM.  This requires that the HII bubble probes sufficiently far from the central system where only $\sim1$ photon per baryon is required to keep the IGM ionised.  Internally to the HII bubble, the recombination rates are high enough such that they require more than one ionising photon per baryon to maintain ionisation down to $z=6$.  This means that the sources that ionise the gas close to the galaxy will begin to mix as the gas recombines and is reionized by different sources.  The cyan regions indicate gas that has been ionised by photons that have been emitted from both middle aged and old stars.

By $z=15.35$, $27\,{\rm Myr}$ after the previous snapshot, the HII bubble has now expanded to more than 1cMpc in diameter.  Much more star formation has occurred and much of the gas has recombined and been reionized so that the shell structure is starting to disappear.  This process will continue as more stellar populations form and ionize the recombining gas.

This behaviour is not unique to stellar age and we find that this same shell structure\footnote{Note that the exact width of the shells is sensitive to how we discretise the different source populations into tracer bins.  For stellar age, we expect the HII bubbles to exhibit smooth gradients of gas ionised by different age stars within a given shell.  The discontinuities will occur when a new generation of stars form and thus our bins approximate this behaviour.} also appears when tracing stellar metallicity.  In Figure~\ref{metal}, we plot a snapshot of the same halo at $z=15.67$ that shows density, temperature, HII fraction, and HII fraction coloured by the different stellar metallicity tracers\footnote{Note that the system has been rotated in this figure compared to Figure~\ref{onion} to better highlight the shell structure.}.  The most massive galaxy is centred in the image.  From the temperature map, it is evident that multiple supernovae have exploded in the central regions of the most massive galaxy, heating the gas to temperatures above $10^5$K.  The structure of the HII bubble is very prominent in the temperature map and gas at $T>10^4$K can be seen outlining the HII bubble.  From the third panel in Figure~\ref{metal}, it is evident that not all gas within the HII bubble is ionised to the same level and much more of the mass in the top lobe is ionised compared to the bottom lobe.  Most interesting is the structure in the tracer map where there is a thin green region of gas ionised by intermediate metallicity stars surrounding a double-lobed red HII bubble that was ionised by the lowest metallicity stars.  Once the first generations of stars age and eject metals into the surrounding gas, a more metal rich generation of stars can form and emit ionising photons.  Thus the delayed onset of this second generation of more metal rich stars leads to a similar shell structure that we have seen when tracing stellar age.  However, this occurs much later when tracing metallicity as it takes longer for the metal rich generation of stars to form than it does for the stars to age.  

\begin{figure*}
\centerline{\includegraphics[scale=1,trim={0 0 0 0},clip]{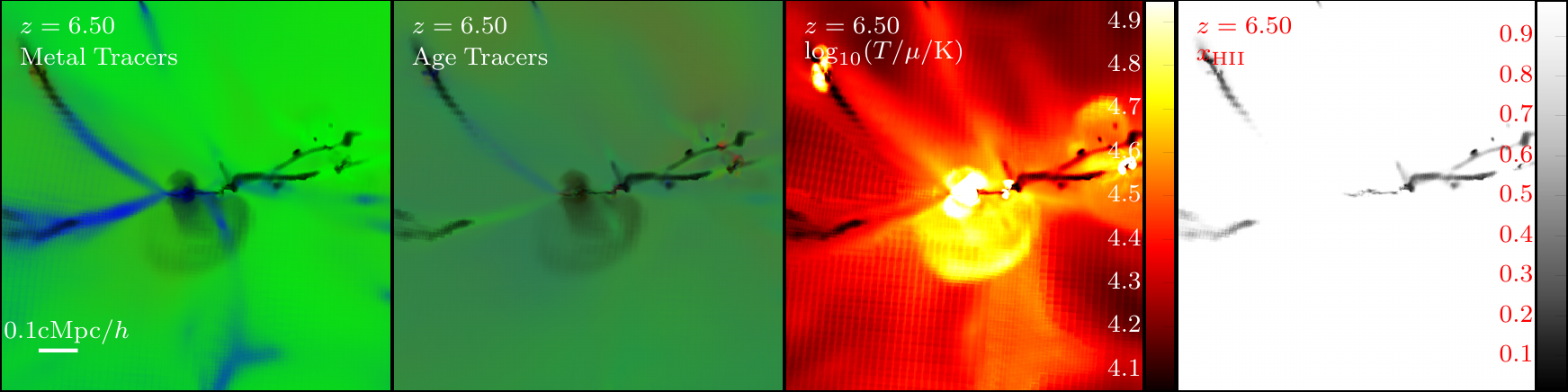}}
\caption{Stellar metallicity tracer map, stellar age tracer map, temperature map, and HII fraction map at $z=6.50$ in a $1$cMpc$h^{-1}$ box around the most massive galaxy in the simulation are shown in the first, second, third, and fourth panels, respectively.  Quantities represent the mass-weighted mean along the axis.  The low density IGM is ionised by different sources compared to the filaments.  Furthermore, although the two filaments feeding the galaxy from the left have been ionised by the same metallicity stars, they have been ionised by different aged stars.  There is clear evidence for collisional ionisation in the dark plume beneath the galaxy in the first and second panels.  This gas appears dark in the tracer map but is clearly ionised in the fourth panel indicating that collisional ionisation is dominant in this region.}
\label{filament}
\end{figure*}

Since the timing of ionisation affects the gas temperature (it cools due to expansion), the green shell of gas that was ionised by intermediate metallicity stars has a temperature that is $\sim2-3\times$ higher than in the other parts of the HII bubble that were ionised by the lowest metallicity stars.  Here, the difference in temperature is entirely due to differences in the ionisation fraction and timing of ionisation.  More generally, different sources (e.g., stars versus AGN) will have different SEDs that can provide different amounts of photoheating.  The photon tracer algorithm is ideal for isolating this kind of effect.

In Figure~\ref{mass_filament}, we plot simulation results at $z=11.65$ and show density, temperature, HII fraction, and HII fraction coloured by the different halo mass tracers around the most massive galaxy.  Most of the right panel appears blue, indicating that the majority of this region has been ionised by UV photons that were emitted by a halo with $M>10^{10}$M$_{\odot}h^{-1}$.  This galaxy is being fed by a dense, mostly neutral filament (see the first and third panels) and is surrounded by supernova-heated gas (second panel).  There are two green lobes on either side of the main filaments in the fourth panel that are feeding the central system, demonstrating that these regions were ionised at a previous point in time when the galaxy was less massive.  This galaxy is not isolated.  Most of the filaments in the image have remained neutral; however, galaxies that have formed along the filaments that are forming stars can ionise some of the high density gas.  Multiple red and green regions appear along the filaments, illuminating the locations of satellite galaxies.  These systems all live in the HII bubble of the most massive galaxy.

The first and second panels of Figure~\ref{filament} show a snapshot at $z=6.50$ where the ionised regions have been coloured based on the metallicity of the stars or the age of the stars that provided the ionising photons, respectively.  These images show that the filaments are ionised by different sources compared to the surrounding IGM.  In the first panel, the two prominent filaments feeding the galaxy from the left appear blue, indicating ionisation by high metallicity stars.  In the second panel, the top filament is slightly blue while the bottom filament is more green, showing that different aged stars have ionised each filament.  Although similar metallicity stars ionised these filaments, the filaments were ionised by different aged stellar populations in the galaxy.  Tracing ionisation based on stellar metallicity is a better indicator of the timing of ionisation compared to stellar age since the metallicity of the stellar populations in galaxies tends to increase with time.  Since these filaments appear blue, they were clearly ionised more recently compared to the surrounding IGM.

Finally, we note that there is a dark plume of gas below the galaxy in the first and second panels of Figure~\ref{filament}.  This gas appears hot in the temperature map (third panel) and ionised in the HII map (fourth panel).  This gas has been ejected from the system via supernova and represents regions where collisional ionisation is the dominant mode of ionisation.  This shows that collisional ionisation is important in and around galaxies and being able to differentiate between the gas that has been collisionally ionised versus photoionised is important for interpreting observations of Ly$\alpha$ emission \citep[e.g.][]{Dijkstra2017}.

\subsection{The contribution of different classes of sources}
\subsubsection{Reionization by volume}
\begin{figure*}
\centerline{\includegraphics[scale=1,trim={0 1cm 0 1cm},clip]{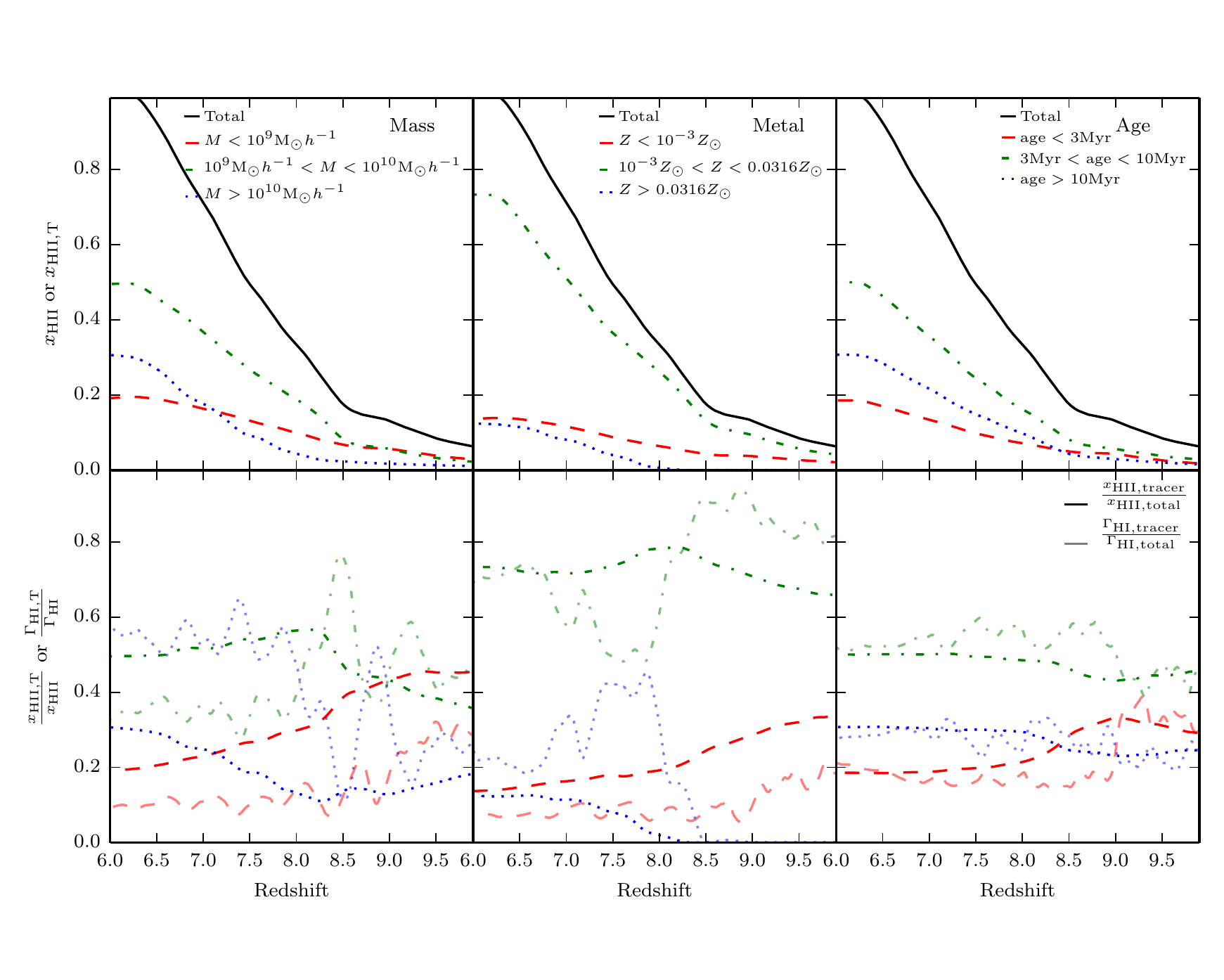}}
\caption{{\it Top}. Evolution of the volume filling fraction of HII ($x_{\rm HII}$ for total or $x_{\rm HII,T}$ for a tracer group) as a function of redshift for the halo mass tracers (left), stellar metallicity tracers (centre), and stellar age tracers (right).  The black line shows the total volume filling fraction of ionised gas while the red, green, and blue lines show the volume filling fraction of ionised gas which was ionised by photons in each of the different tracer bins T1, T2, and T3, respectively. {\it Bottom.} Fractional contribution to the total volume filling fraction of ionised gas for each of the different tracers are shown as the darker lines.  The lighter lines represent the fractional contribution of each of the different source populations to the volume-weighted mean photoionisation rate of HI in ionised regions (defined as cells with $x_{\rm HII}>0.5$).  Especially when tracing by halo mass, it is not necessarily the case that the sources that dominate the meta-galactic UV background at $z=6$ are responsible for ionising the majority of the gas in the Universe.}
\label{VWion}
\end{figure*}

Using the photon tracers, we can average over the entire simulation volume or individual environments in the simulation at all redshifts to better understand which sources are driving reionization.  We begin this analysis by measuring the contribution of each source population to the volume-weighted ionised fraction of gas ($x_{\rm HII}$) as a function of redshift.  In the top row of Figure~\ref{VWion}, we plot the redshift evolution of this quantity for each of the three simulations that trace halo mass, stellar metallicity, and stellar age.

Beginning with halo mass, in the top left panel of Figure~\ref{VWion}, we show the total volume-weighted ionisation fraction as a function of redshift (black line, $x_{\rm HII}$) and the fraction of gas ionised by different halo masses as a function of redshift (coloured lines, $x_{\rm HII,T}$).  By $z=6$, at the end of reionization, the contribution from the lowest mass, intermediate mass, and highest mass systems to the volume-weighted ionisation fraction is $\sim20\%$, $50\%$, and $30\%$, respectively.  Thus, in this simulation, the intermediate mass systems with $10^9{\rm M_{\odot}}h^{-1}<M<10^{10}{\rm M_{\odot}}h^{-1}$ ionize most of the gas.  Although each halo mass bin contributes significantly to reionization, the time period over which each halo mass bin ionises their respective volumes is different.  For example, of the volume of gas that is ionised by low mass haloes by $z=6$, 50\% was ionised at $z>8.07$.  This quantity decreases with increasing halo mass and is $z>7.62$ for intermediate mass haloes and $z>7.15$ for high mass haloes.  Thus the low mass haloes ionize much of the volume first, followed by the intermediate mass haloes, with a smaller contribution from the high mass haloes towards the end of reionization, as expected from the growth in the star formation rates as a function of time for different mass haloes \citep{Behroozi2018,Moster2018}.  

In the bottom left panel of Figure~\ref{VWion}, we show the fractional contribution of each halo mass tracer bin to the total ionisation fraction as a function of redshift (i.e., $x_{\rm HII,T}/x_{\rm HII,total}$).  At $z=10$, when the simulation is $\lesssim10\%$ ionised, the contributions from the intermediate and lowest mass systems dominate while the highest mass system only contributes $\sim20\%$ to the total ionisation fraction.  At this redshift, there is only one system in the simulation that belongs to the highest mass halo bin and while it is efficient at ionising its surroundings, the contribution to the total volume-weighted ionisation fraction remains low.  The lowest and intermediate mass systems maintain a roughly equal contribution to the total ionisation fraction until $z\sim8.5$.  The contribution from the highest mass systems overtakes the contribution from the lowest mass systems at $z\sim7.2$. 

At $z<10$, the contribution to the volume-weighted ionisation fraction from each tracer group, $x_{\rm HII,T}$, increases continuously.  By this point in the simulation, the ionisation fronts are beginning to expand into the lower density IGM.  In these gas phases, the recombination time is long and thus few photons per baryon are required to maintain ionisation down to $z=6$.  Since the IGM dominates the Universe by volume, $x_{\rm HII,T}$ is unlikely to decrease by any significant amount once the ionisation fronts have fully expanded into the IGM.

Although the intermediate mass systems are responsible for reionizing most of the volume by $z=6$, this does not necessarily mean that they dominate the volume-weighted photoionisation rate of neutral hydrogen, $\Gamma_{\rm HI}$.  The contribution of each different tracer group to $\Gamma_{\rm HI}$ was studied in detail in \cite{Katz2018} and in the bottom row of Figure~\ref{VWion}, we compare the fractional contribution of each tracer group to $\Gamma_{\rm HI}$ with the fractional contribution to the volume-weighted ionisation fraction.  Most notably, while the ionisation fraction at $z=6$ is dominated by the intermediate mass haloes, $\Gamma_{\rm HI}$ is dominated by the most massive systems.  {\it In other words, the systems that are responsible for ionising the gas during reionization are not necessarily the same systems that dominate the UV background at a given redshift, especially in the post reionization epoch}.  Once a region is ionised, photons from other sources can fill into that region and contribute to the photoionisation rate.  This is particularly important after reionization as $\Gamma_{\rm HI}$ will continue to increase until $z\sim4$ \citep{Bolton2007,Wyithe2011,Calverley2011,Becker2013} after the ionisation fraction has been saturated.  

The redshift evolution of the contribution of the different mass galaxies to the volume-weighted ionisation fraction is smoother than the evolution of each tracer group's contribution to $\Gamma_{\rm HI}$.  $\Gamma_{\rm HI}$ is much more sensitive to the instantaneous star formation rate of each tracer group.  At high redshift, when a small fraction of the volume of the simulation is ionised, $\Gamma_{\rm HI}$ relies on a small patch of the universe where the star formation history is bursty, hence the wiggles in $\Gamma_{\rm HI}$ as a function of redshift.  However, we do not see this behaviour in the volume-weighted ionised fraction because regardless of the number of ionising photons in a cell, the ionisation fraction will saturate at 1.  It then takes time for these photons to penetrate out into the IGM and ionize the gas.  Thus, at $z<10$, we do not see strong changes in $x_{\rm HII}$. 

\begin{figure*}
\centerline{\includegraphics[scale=1,trim={0 1cm 0 1cm},clip]{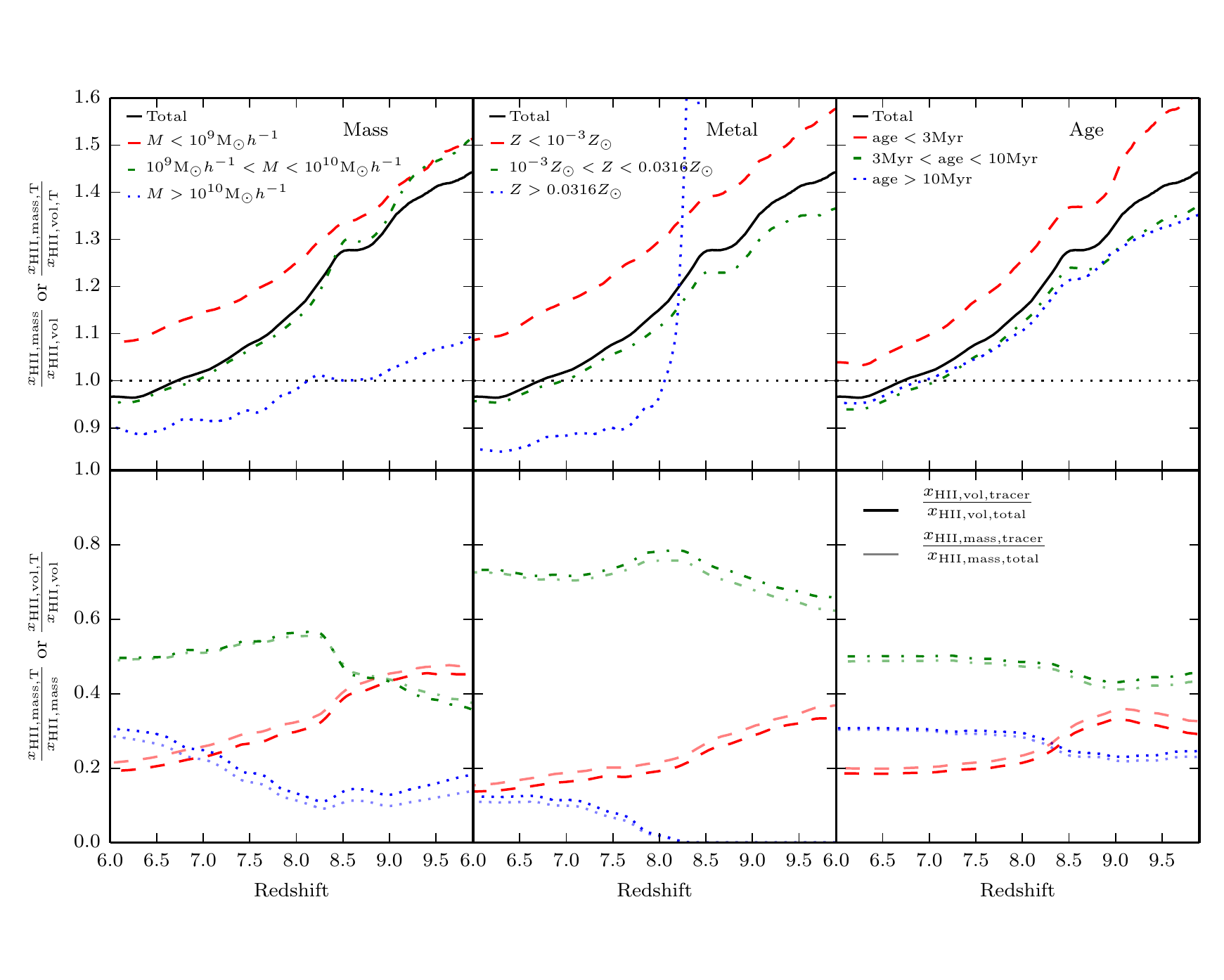}}
\caption{{\it Top}. Evolution of the ratio of the mass-weighted ionisation fraction to the volume-weighted ionisation fraction as a function of redshift for the halo mass tracers (left), stellar metallicity tracers (centre), and stellar age tracers (right).  The black line shows the total ratio while the red, green, and blue lines show this ratio split for the different tracer groups T1, T2, and T3, respectively.  At high redshift, this ratio is high as only the dense regions around galaxies are ionised.  As the HII bubbles penetrate into the voids, this ratio decreases and drops below 1 as the filaments and galaxies are the last regions to be ionised.  The ratio is different depending on the source.  {\it Bottom}.  Comparison between the fractional contribution to the volume-weighted ionisation fraction (darker) and mass-weighted ionisation fraction (lighter) as a function of redshift.  Generally these two quantities retain a similar value; however, there are small deviations that show that different types of sources are more responsible for ionising the mass in the universe compared to the volume of the Universe and vice versa.}
\label{MWion}
\end{figure*}

When tracing stellar metallicity, in the top centre panel of Figure~\ref{VWion}, by $z=6$ the intermediate metallicity stars are most dominant in their contribution to the volume-weighted ionisation fraction ($\sim 70\%$ of the gas by volume).  Stars with the lowest metallicity contribute just over 10\% to $x_{\rm HII}$, which is approximately the same as the highest metallicity stars.  We see a similar effect in the lower middle panel of Figure~\ref{VWion} as when tracing halo mass: the contribution of the different sources to the volume-weighted photoionisation rate may be different to the contribution to $x_{\rm HII}$.  In this case, the fractional contribution of the intermediate metallicity stars to $\Gamma_{\rm HI}$ is similar to their contribution to $x_{\rm HII}$; however, the highest metallicity stars contribute more to $\Gamma_{\rm HI}$ compared to $x_{\rm HII}$, while the opposite is true for the lowest metallicity stars.  The fractional contribution of the lowest metallicity stars to $x_{\rm HII}$ decreases as a function of redshift as the systems become more metal enriched (see also \citealt[e.g.][]{Paar2013,Xu2016,Kimm2017} who similarly find a decreasing contribution of Pop.~III stars with decreasing redshift).  As these stars age, they emit fewer ionising photons and thus, when absorbed, the contribution to $\Gamma_{\rm HI}$ will decrease.  Hence, it is not particularly surprising that there is a mismatch between the contribution to $\Gamma_{\rm HI}$ and $x_{\rm HII}$.

A different behaviour is seen when tracing stellar age.  From the top right panel of Figure~\ref{VWion}, it is evident that the middle aged stars are responsible for reionizing most of the simulation by $z=6$.  Since younger stars are preferentially found in very dense regions with high recombination rates, little of their radiation escapes since this radiation is emitted before SN explode \citep{Kimm2017,Trebitsch2017,Rosdahl2018}.  Surprisingly, by $z=6$ the oldest stars have ionised more of the box by volume compared to the youngest stars despite emitting far fewer ionising photons (by more than a factor of 6).  This is a direct result of the different escape fractions of LyC photons for each of the different stellar ages \citep{Rosdahl2018}.  In \cite{Katz2018}, we found an average escape fraction of $\sim45\%$ for the old stars, while this quantity is only $\sim5\%$ for the young stars. Compared to the fractional contribution to the volume-weighted photoionisation rate, we see from the bottom right panel of Figure~\ref{VWion} that when tracing stellar age, the fractional contribution to the volume-weighted ionised fraction is nearly the same.  Thus in this case, the sources which ionised the gas contribute equally to the UV background.

\subsubsection{Reionization by mass}

Comparing the mass-weighted ionisation fraction ($x_{\rm HII,mass}$) to the volume-weighted ionisation fraction ($x_{\rm HII,vol}$) as a function of redshift can provide insight into how different regions of the Universe are being ionised.  In the top row of Figure~\ref{MWion}, we plot the ratio of these two quantities as a function of redshift for the entire simulation as well as for each of the different tracers.  Our simulations are consistent with a reionization topology where the filaments ionize last as they remain self-shielded while the surrounding IGM is ionised \citep{Gnedin2000a,Finlator2009,Choudhury2009}.  

Using the photon absorption tracers, we can now measure this fraction individually for each source population.  In the top left panel of Figure~\ref{MWion}, we show how $x_{\rm HII,mass}/x_{\rm HII,vol}$ varies for gas ionised by different mass haloes.  The green line follows the black line indicating that the intermediate mass systems follow the globally averaged quantity.  These systems form early on and dominate $x_{\rm HII,vol}$ at $z\lesssim9$.  Because these systems form early, much of the CGM is still neutral and thus this class of sources must be responsible for ionising different gas densities.  By the time these systems grow in mass and move from the intermediate mass bin to the high mass bin, much of the surrounding CGM has been ionised and thus, photons from sources in the highest mass bin, T3, are expected to reach lower density regions sooner.  The blue line in the top left panel of Figure~\ref{MWion} represents $x_{\rm HII,mass}/x_{\rm HII,vol}$ for the highest mass systems and this ratio drops below one at much higher redshift compared to the globally averaged quantity, which demonstrates this effect.  In contrast, $x_{\rm HII,mass}/x_{\rm HII,vol}$ never drops below one for the lowest mass systems.  This shows that the HII regions from these systems rarely extend far away from their host galaxies and into mean density regions.

Performing a similar experiment with the stellar metallicity tracers, we see similar behaviour.  The intermediate metallicity stars that form early and ionize most of the volume of the simulation track the globally averaged quantity reasonably well.  Similarly, the highest metallicity systems that form very late and in large, pre-existing HII bubbles have $x_{\rm HII,mass}/x_{\rm HII,vol}$ that drops below one much earlier than the global quantity.  At $z\sim8.2$, the blue line in the top centre panel of Figure~\ref{MWion} increases rapidly towards very high values as the redshift increases.  These stars form in very dense regions with high recombination rates and thus they must ionize their birth cloud before the photons can escape the galaxy.  Hence a very high ratio of $x_{\rm HII,mass}/x_{\rm HII,vol}$.  However, this period is short lived because once the birth cloud has been ionised, the photons penetrate far away from the galaxy since they exist in a large HII bubble.  Similar to tracing halo mass, the lowest metallicity stars also have $x_{\rm HII,mass}/x_{\rm HII,vol}>1$ for all redshifts. 

Tracing by stellar age illuminates a different effect.  In the top right panel of Figure~\ref{MWion}, we show how $x_{\rm HII,mass}/x_{\rm HII,vol}$ is split for the different age tracers.  The red line, which represents the young stars, always has a ratio greater than one as well as the globally averaged quantity (the black line).  This is expected because young stars form in dense neutral regions which require multiple photons per baryon to maintain the ionisation state and thus this population of stars has a lower escape fraction compared to the older generations.  Clearly, not all of the photons from the young stars are absorbed locally because the ratio does decrease with redshift indicating that the young stars are ionising some low density regions.  In contrast, the middle and old age stars have ratios that remain below the globally averaged quantity because the young stars have done most of the work in ionising the high density regions.  

In the bottom row of Figure~\ref{MWion}, we compare the fractional contribution of each of the different tracers to $x_{\rm HII,mass}$ with the fractional contribution to $x_{\rm HII,vol}$.  When the fractional contributions are the same, we expect that the source will follow the globally averaged quantity in the top row of plots.  When the contribution to $x_{\rm HII,mass}$ is larger than the contribution to $x_{\rm HII,vol}$, we expect the system to fall above the globally averaged quantity in the top row.  Finally, if the contribution to $x_{\rm HII,mass}$ is smaller than the contribution to $x_{\rm HII,vol}$, we expect that the corresponding line in the top row will fall below the globally averaged ratio of $x_{\rm HII,mass}/x_{\rm HII,vol}$.  In the bottom left panel of Figure~\ref{MWion}, the fractional contribution of the high mass systems to $x_{\rm HII,vol}$ is always larger than to $x_{\rm HII,mass}$.  By $z=6$, the high mass haloes have fractionally contributed 2\% more to ionising the volume than the mass.  While this fractional difference seems small, only $96\%$ of the total mass in the box is ionised and thus these high mass haloes actually ionize 10\% more volume than mass.  The opposite is true for the low mass haloes.  By $z=6$, the low mass haloes have fractionally contributed 2\% more to ionising the mass rather than the volume which leads to an 8\% difference in the total mass ionised compared to the total volume. 

\subsubsection{Reionization as a function of over-density}
\begin{figure*}
\centerline{\includegraphics[scale=1,trim={0 1.5cm 0 0cm},clip]{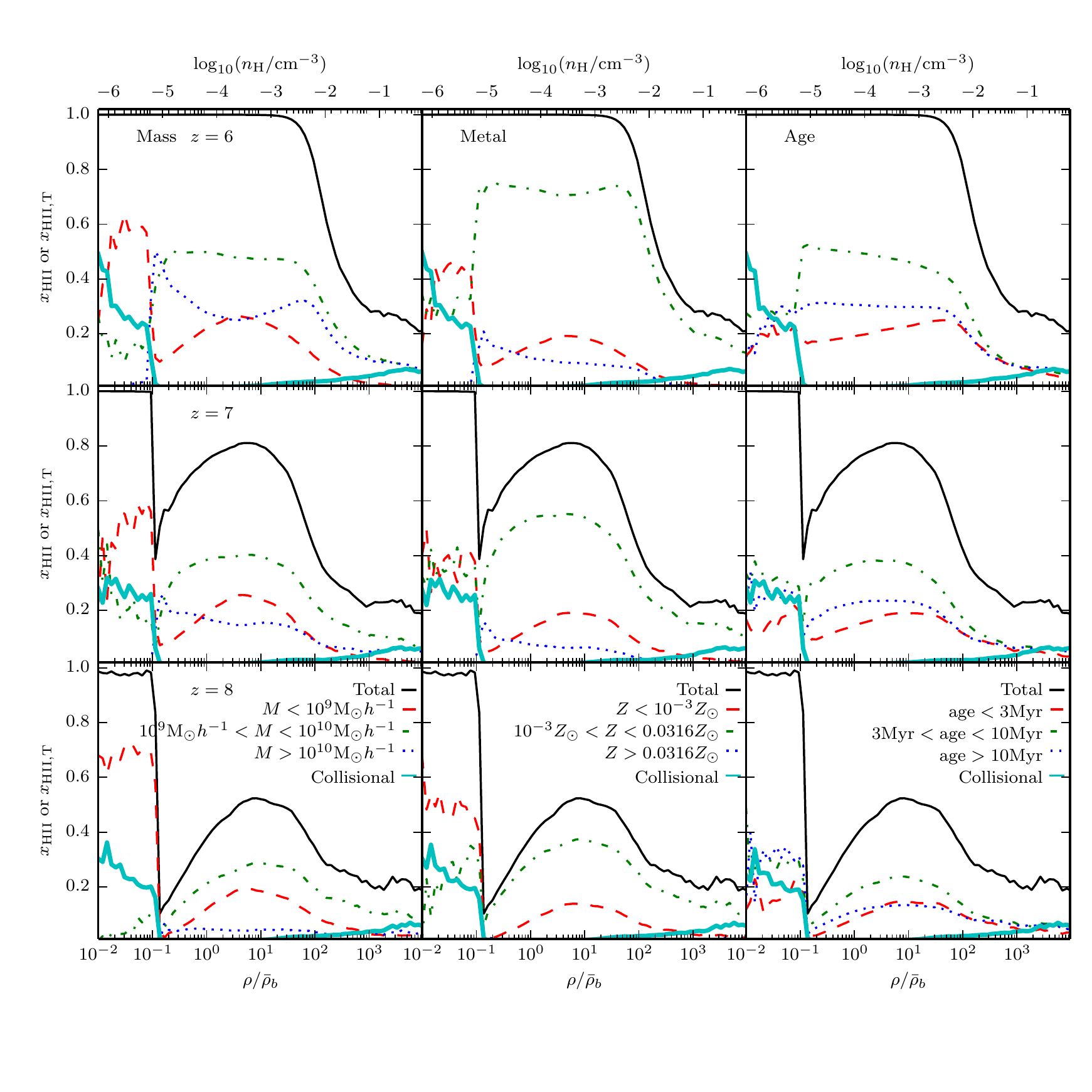}}
\caption{Volume-weighted ionisation fraction as a function of over-density at $z=6$ (top row), $z=7$ (middle row), and $z=8$ (bottom row).  The results for the halo mass, stellar metallicity, and stellar age tracers are shown in the left, centre, and right columns, respectively.  The gas that has been ionised by sources in T1, T2, and T3, are shown in red, green, and blue.  Gas that is collisionally ionised is shown in cyan.  Collisional ionisation becomes important in supernova heated gas at $\rho/\bar{\rho}_b<10^{-1}$ as well as at the highest over densities, $\rho/\bar{\rho}_b>10^3$, inside of galaxies.  At $z=6$, the box is completely ionised and the turnover at $\rho/\bar{\rho}_b\gtrsim30$ corresponds to self-shielded gas.  At higher redshifts much of the IGM is still neutral, hence there is an additional turnover at $\rho/\bar{\rho}_b\lesssim5-10$.  In all cases, most of the gas at intermediate over densities has been ionised by sources in T2.  Note that the density labels on the top axis only apply to the $z=6$ plots.}
\label{rhogas}
\end{figure*}

Rather than studying the general contribution of each class of sources to either the mass- or volume-weighted ionisation fraction, at individual redshifts, we can quantify the contribution of each source to the ionisation fraction at a specific gas density.  In Figure~\ref{rhogas}, we plot the ionisation fraction as a function of $\rho/\bar{\rho}_b$ at $z=6$ (top row), $z=7$ (middle row), and $z=8$ (bottom row), where $\bar{\rho}$ is the mean baryon density at that redshift.  By $z=6$, over-densities $\lesssim100$ are nearly completely ionised while at over-densities higher than this value, the ionisation fraction drops off steeply.  This is the regime where self-shielding occurs and a significant portion of the gas remains neutral.  Moving to $z=7$ and $z=8$, it is no longer the case that all over-densities $\lesssim100$ are completely ionised.  At $\rho/\bar{\rho}_b\lesssim0.1$, we observe an ionisation fraction of $\sim100\%$.  This is the gas that has been affected by supernova feedback and thus it remains close to the galaxy where the UV flux is still very strong.  However, there is a very quick drop off in the ionisation fraction at slightly higher over-densities which is the transition to the IGM.  At these redshifts, the ionisation fronts have not completely penetrated the lowest density voids and hence the ionisation fraction remains low.  The ionisation fraction continues to increase moving towards higher over-densities as we begin to probe the CGM around galaxies.  The ionisation fraction peaks at $\rho/\bar{\rho}_b\sim10$ at both $z=7$ and $z=8$ before decreasing again due to self-shielding.

\begin{figure*}
\centerline{\includegraphics[scale=1,trim={0 0cm 0 0cm},clip]{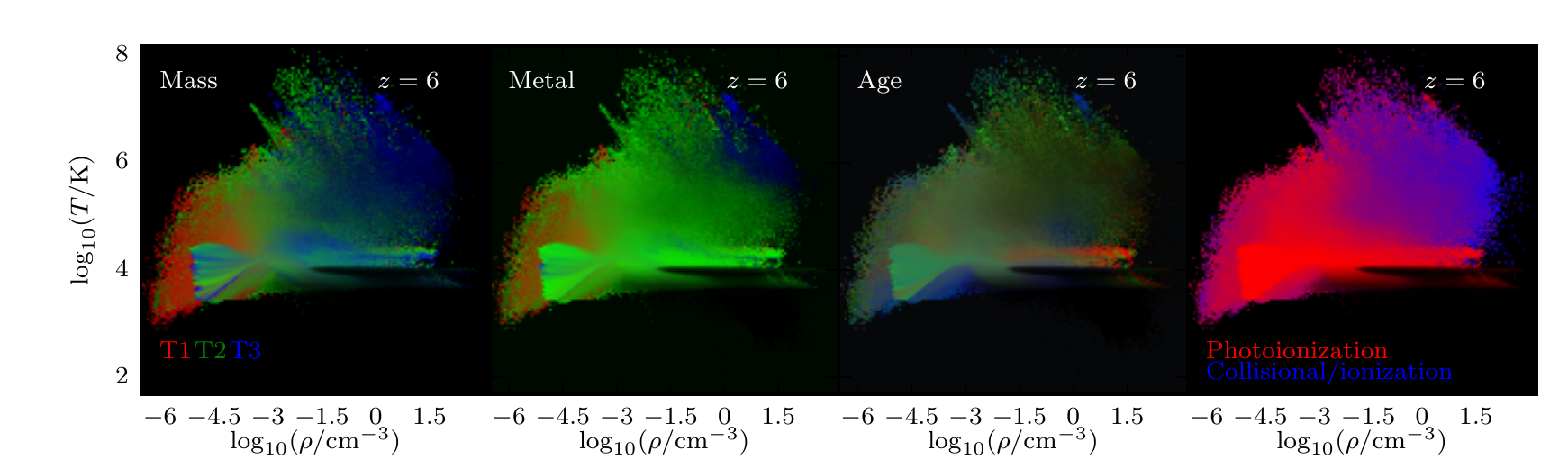}}
\caption{Phase-space diagrams of density versus temperature for ionised gas in the simulation at $z=6$.  In the left three panels, the fraction of the mass in each bin that has been ionised by sources in T1, T2, and T3 are shown in red, green, and blue, respectively.  The first, second, and third panels represent the three simulations tracing halo mass, stellar metallicity, and stellar age, respectively.  In the fourth panel, the contribution from collisional ionisation is shown in blue compared to photoionization in red.  Collisional ionisation is most important in the supernova heated gas.}
\label{psgas}
\end{figure*}

We can now split the ionisation fraction into the contribution from each different type of source.  Beginning with halo mass (left column of Figure~\ref{rhogas}), the intermediate mass haloes are responsible for ionising most of the gas at mean density across all three redshifts.  More generally, the intermediate mass systems dominate the ionisation fraction of all over-densities $>0.1$ at $z\leq8$.  At the lowest over-densities with $\rho/\bar{\rho}_b<0.1$, the lowest mass systems have the most important contribution at all redshifts while the contribution of this tracer group to the highest densities ($\rho/\bar{\rho}_b>10^3$) is effectively negligible.  Most of the very dense gas exists in the more massive objects and the recombination rate is so high that the weaker radiation fields in the smaller systems are not able to effectively ionize this gas.  In contrast, supernova feedback is more effective in the lower mass systems \citep{Dekel1986} and thus most of the volume of the lowest density gas exists around the lowest mass haloes.  The highest mass systems only have an impact at $z<8$ after a significant number of these systems have formed which explains why the contribution of the highest mass systems at all over-densities grows with decreasing redshift.

Comparing the three profiles at $z=6$, the shapes of the contribution to the ionisation fraction as a function of over-density differ for low, intermediate, and high mass haloes.  The shape of the green line, which represents intermediate mass systems, mimics that of the global line indicating that these systems are responsible for photoionizing all different types of gas phases.  Since the lowest mass systems only ionise surrounding gas, their profile in Figure~\ref{rhogas} only matches the global profile at $z=7$ and $z=8$.  At $z=6$, the high mass haloes, represented by the blue line, have an inverted profile compared to the lowest mass haloes.  There is a peak at moderately high densities ($\rho/\bar{\rho}_b\sim10^2$), which is the regime where recombinations become important and a continuous source of ionising photons is required to maintain the ionisation state of the gas.  The contribution decreases moving towards mean densities, while having a peak in slightly under-dense regions.  At $0.1\lesssim\rho/\bar{\rho}_b\lesssim0.2$, the contribution from the highest mass haloes is dominant over both the intermediate and low mass systems indicating that the highest mass systems complete the ionisation of the voids by $z=6$, consistent with the results in Figure~\ref{MWion}.

We should emphasise that while very close to 100\% of the volume and mass in the box is ionised by photoionisation, collisional ionisation can play a significant role in certain regions of parameter space.  When the ionisation fractions from the tracer groups do not sum to the total ionisation fraction in the cell, the remaining fraction is due to collisional ionisation (see \citealt{Katz2018}).  In Figure~\ref{rhogas}, we plot the fraction of gas that has been collisionally ionised as a function of over-density in cyan.  At $\rho/\bar{\rho}_b<0.1$, which is the regime where supernova have heated the gas, collisional ionisation is clearly non-negligible and is responsible for $\sim20\%-50\%$ of the total ionisation fraction at these densities.  The equation for the collisional ionisation rate due to collisions with electrons in units of cm$^{3}$s$^{-1}$ is
\begin{equation}
\label{cion}
C_{{\rm HI,}e}(T,n_e)=\frac{5.85\times10^{-11}\sqrt{T}}{1+\sqrt{\frac{T}{10^5}}}e^{\frac{-157809.1}{T}}n_e\ {\rm [cm^3s^{-1}]},
\end{equation}
where $T$ is the temperature in K and $n_e$ is the electron number density in cm$^{-3}$.  Collisional ionisation becomes most important in gas with high temperatures and also at high electron number densities and this is evident in Figure~\ref{rhogas}.

This can be seen more clearly in Figure~\ref{psgas}, where we plot the temperature-density phase-space diagram for ionised gas in the simulation at $z=6$.  In the left three panels, we have coloured the pixels based on which type of source was responsible for ionising the gas while in the right panel, the diagram is coloured depending on whether the gas was collisionally or photoionized.  In the right panel of Figure~\ref{psgas}, at the very highest densities and highest temperatures, we see many blue pixels indicating that collisional ionisation is the dominant mode of ionisation in these regions of phase-space.  At the very lowest densities, the colour appears purple which is gas that was ionised by a mix of photoionization and collisional ionisation.  Much of this gas was likely affected by a supernova and has gone on to expand and cool.  The exact temperature-density phase-space where collisional ionization is likely to be important is given by Equation~\ref{cion}; however, the novelty of our photon tracer algorithm is that we can directly measure the exact fractional contribution of collisional ionization versus photoionization for each gas cell in the simulation.

The behaviour of the stellar metallicity tracers is very similar to the halo mass tracers as a function of gas over-density, with the exception that the intermediate metallicity stars are more dominant compared to the intermediate mass galaxies.  However, when analysing the stellar age tracers, we observe different behaviour.  At $z=6$, in the top right panel of Figure~\ref{rhogas}, the contribution decreases and remains constant with over-density for the middle aged and old stars, respectively. In contrast, the ionisation fraction due to the youngest stars increases with over-density.  In the third panel of Figure~\ref{psgas}, we plot the temperature-density phase-space diagram for the ionised gas at $z=6$ coloured by which age stars ionised the region.  There is a prominent red streak extending from $-1.5\lesssim\log_{10}(\rho/{\rm cm^{-3}})\lesssim1.5$ at a temperature of $\sim10^4$K indicating that the youngest stars are responsible for ionising much of the gas that is falling into the galaxy and creating stars, consistent with Figure~\ref{rhogas}.  Gas at this temperature that is present at lower densities appears blue-green in this image, demonstrating that the middle aged stars are responsible for ionising much of the CGM.

In the left three panels of Figure~\ref{psgas}, we find that there are clear striations in the temperature density relation at $\rho\sim10^{-4.5}$cm$^{-3}$.  This scatter is a result of gas at this density being ionised at different times (see e.g., \citealt{Furlanetto2009}).  The coloured striations are then a result of bursts of star formation that release large quantities of ionising photons in different tracer groups and reionize the IGM at different times.

\section{Caveats}
\label{cavs}
The work presented here demonstrates the ability of our recently developed photon tracer algorithm to unravel the complex nature of the escape of ionising photons during reionization.  There are, however, various caveats that should be kept in mind, many of which are also listed in \cite{Katz2018}.  As we discussed earlier, the limited box size and spatial resolution as well as finite mass resolution of our simulations prohibit us from making conclusive predictions for exactly which mass haloes, metallicity stars, and age stars reionized the real Universe.  In order to address these questions with more confidence, the ionising sources in the multiphase ISM must be simulated with sufficient resolution to obtain converged results for the escape of LyC photons.  This must be done for the wide mass range of haloes hosting ionising sources that potentially contribute significantly to reionization.  

Additionally, our current simulations still employ rather simplistic sub-grid models for star formation and stellar feedback.  Our delayed cooling stellar feedback places gas in an unstable region of temperature-density phase-space.  Furthermore, stellar populations are expected to have a delay-time distribution for supernova explosions whereas here we model them as occurring $10\,{\rm Myr}$ after formation.  Finally, stars in a cluster have a small spread in their formation times which may impact our conclusions on the contribution of different aged stars to reionization.

Our work here makes use of the M1 closure which prevents photon beams of equal and opposite flux from mixing.  This can become particularly problematic in the optically thin regime, when much of the simulation volume is ionised and might prevent oppositely traveling photon fronts from crossing one another \citep{Rosdahl2013}.  Most of our analysis in this work is not done in this regime, hence we expect that the lack of mixing due to the M1 closure will have only a small impact on which sources are ionising the simulated volume.    

With these caveats in mind, we expect that many of our conclusions will still hold.  In particular, our findings that different classes of sources ionise different densities is very likely robust. The relevant processes that drive this effect, such as the clustering of different mass galaxies, collapse times of haloes of different masses, and star formation histories dependent on galaxy mass, are well captured by our simulations. Our algorithm is also well suited for the higher dynamic range reionization simulations that will hopefully overcome many of the limitations of our current simulations as the added computational cost of the tracer algorithm is moderate.

\section{Discussion \& Conclusions}
\label{discussion}
We have used here the photon tracer algorithm first presented in \cite{Katz2018} to measure the contribution of various classes of ionising sources to the reionization of the Universe in simulations with the cosmological multifrequency radiation hydrodynamics code {\small RAMSES-RT}.  Tracing the sources of ionising photons has allowed us to measure the redshift evolution of the ionised fraction (by mass and volume) due to photons emitted by stars with different metallicities and ages and hosted by different mass haloes.  Our results can be summarised as follows:

\begin{enumerate}
\item  The early growth of HII bubbles proceeds in concentric shells that reflect variations in the star formation history of the galaxies acting as the sources of reionization.  These variations lead to temperature differences in the radial direction along the bubbles. As the bubbles grow, recombinations become more important and the shell like structure of the HII bubbles is erased after a few generations of star formation.

\item By $z=6$, both the mass and volume in our simulations are predominantly reionized by stars located in intermediate mass haloes with $10^9{\rm M_{\odot}}h^{-1}<M<10^{10}{\rm M_{\odot}}h^{-1}$,  stars with intermediate metallicity, $10^{-3}Z_{\odot}<Z<10^{-1.5}Z_{\odot}$, and intermediate age stars with $3\,{\rm Myr} < t < 10 \, {\rm Myr}$. 

\item The sources that provide the majority of the photons that (re-)ionize the IGM are not necessarily the same sources that provide the majority of photons that make up the meta-galactic UV background.  This is the case if source categories are defined both by host halo mass and by stellar metallicity and is the result of the different formation times of the stars in the different source categories.

\item Different source categories also dominate the reionization of different gas phases.  For instance, the intermediate mass haloes ionise the majority of the CGM while reionization in the lowest density regions of the IGM is completed by photons emitted in the highest mass host haloes.  Collisional ionisation is significant in both the lowest density supernova heated gas as well as the highest density ionised gas.  Filaments are the last gas phase to be reionized since they typically reside far from galaxies and exhibit high recombination rates.  These regions quickly lose memory of the the sources that first ionised them due to the recombinations.  The sources that maintain the ionisation state of the filaments are thus rather different to those that reionize the IGM surrounding the filaments.

\end{enumerate}

Using our photon tracer and absorption algorithm, we have highlighted the complex interplay of various processes that determine which sources dominate reionization by mass and volume and contribute to the metagalactic UV background.  The results presented here provide strong motivation for employing our algorithm in both larger volumes to capture sources not yet probed by our current simulations (e.g. higher mass galaxies and AGN) and smaller volumes to perform a more detailed study on how radiation escapes from a well resolved ISM.

\section*{Acknowledgements}
We thank the referee for their comments that improved the manuscript.  This work made considerable use of the open source analysis software {\small PYNBODY} \citep{pynbody}.  HK thanks Brasenose College and the support of the Nicholas Kurti Junior Fellowship as well as the Beecroft Fellowship. Support by ERC Advanced Grant 320596 ``The Emergence of Structure during the Epoch of reionization" is gratefully acknowledged.  DS acknowledges support by STFC and ERC Starting Grant 638707 ``Black holes and their host galaxies: coevolution across cosmic time''.  TK acknowledges support by the Yonsei University Future-leading Research Initiative of 2017 (RMS2-2017-22-0150) and in part by the National Research Foundation of Korea (NRF) grant funded by the Korea government (No.2018036146).  JR and JB acknowledge support from the ORAGE project from the Agence Nationale de la Recherche under grand ANR-14-CE33-0016-03.

This work was performed using the DiRAC/Darwin Supercomputer hosted by the University of Cambridge High Performance Computing Service (http://www.hpc.cam.ac.uk/), provided by Dell Inc. using Strategic Research Infrastructure Funding from the Higher Education Funding Council for England and funding from the Science and Technology Facilities Council. 

This work used the DiRAC Complexity system, operated by the University of Leicester IT Services, which forms part of the STFC DiRAC HPC Facility (www.dirac.ac.uk). This equipment is funded by BIS National E-Infrastructure capital grant ST/K000373/1 and  STFC DiRAC Operations grant ST/K0003259/1. 

Furthermore, this work used the DiRAC Data Centric system at Durham University, operated by the Institute for Computational Cosmology on behalf of the STFR DiRAC HPC Facility (www.dirac.ac.uk).  This equipment was funded by the BIS National E-infrastructure capital grant ST/K00042X/1, STFC capital grant ST/K00087X/1, DiRAC operations grant ST/K003267/1 and Durham University.  DiRAC is part of the National E-Infrastructure.

\bibliographystyle{mnras}
\bibliography{refs} 

\bsp	
\label{lastpage}
\end{document}